# Multi-Constrained Topology Optimization via the Topological Sensitivity

Shiguang Deng, Krishnan Suresh *

Mechanical Engineering, University of Wisconsin, Madison, USA

**Abstract**

The objective of this paper is to introduce and demonstrate a robust method for constrained multi-load topology optimization. The method is derived by combining the topological level-set with the classic augmented Lagrangian formulation.

The primary advantages of the proposed method are: (1) it rests on well-established augmented Lagrangian formulation for constrained optimization, (2) the augmented topological level-set can be derived systematically for an arbitrary set of loads and constraints, and (3) the level-set can be updated efficiently. The method is illustrated through numerical experiments.

## 1. INTRODUCTION

Topology optimization has rapidly evolved from an academic exercise into an exciting discipline with numerous industrial applications [1], [2]. Applications include optimization of aircraft components [3], [4], spacecraft modules [5], automobiles components [6], cast components [7], compliant mechanisms [8]–[11], etc.

A typical *single-load* topology optimization problem in structural mechanics may be posed as (see Figure 1):

$$\underset{\Omega \subset D}{Min}\, \varphi(u, \Omega)$$
$$g_i(u, \Omega) \leq 0; i = 1, 2, ..., m$$
$$\text{subject to}$$
$$Ku = f \tag{1.1}$$

where:
- $\varphi$: Objective to be minimized
- $\Omega$: Topology to be computed
- $D$: Domain within which the topology must lie
- $u$: Finite element displacement field
- $K$: Finite element stiffness matrix
- $f$: External force vector
- $g_i$: Constraints
- $m$: Number of constraints

(1.2)

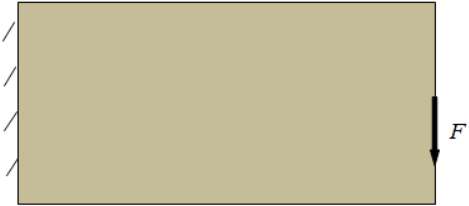

**Figure 1:** A single-load structural problem.

* Corresponding author.
  E-mail address: suresh@engr.wisc.edu

A classic example is compliance minimization:

$$\underset{\Omega \subset D}{Min}\, J = f^T u$$
$$|\Omega| - v_0 \leq 0$$
$$\text{subject to}$$
$$Ku = f \tag{1.3}$$

Similarly, a *stress-constrained* volume-minimization problem [13], [14] (with an additional compliance constraint to avoid pathological situations) may be posed as:

$$\underset{\Omega \subset D}{Min}\, |\Omega|$$
$$\sigma \leq \sigma_{max} \text{ in } \Omega$$
$$J \leq J_{max}$$
$$\text{subject to}$$
$$Ku = f \tag{1.4}$$

where:
- $\sigma$: von Mises Stress
- $\sigma_{max}$: Max. allowable von Mises Stress
- $J$: Compliance
- $J_{max}$: Max. compliance allowed
- $\Omega$: Topology to be computed
- $D$: Region within which the topology must lie

(1.5)

A *multi-load* problem, on the other hand, may be posed as (see Figure 2 for an example of a two-load problem):

$$\underset{\Omega \subset D}{Min}\, \varphi(u_1, u_2, ..., u_N, \Omega)$$
$$g_i(u_1, u_2, ..., u_N, \Omega) \leq 0; i = 1, 2, ..., m$$
$$\text{subject to}$$
$$Ku_n = f_n; n = 1, 2, ..., N \tag{1.6}$$

where:
- $u_n$: Displacement field for load-n
- $f_n$: External force vector for load-n
- $N$: Number of loads

(1.7)

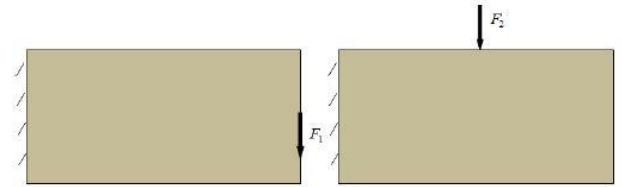

**Figure 2:** A multi-load structural problem.

While various methods have been proposed to solve specific instances of single-load and multi-load problems (see Section 2 for a review), the objective of this paper is to develop a unified method that is applicable to all flavors of single and multi-load problems.

The proposed method relies on the concepts of topological level-set [15]–[19] and augmented Lagrangian [67], and it overcomes the deficiencies of existing methods discussed next.

## 2. LITERATURE REVIEW

Popular strategies for solving constrained topology optimization problems can be classified into two distinct types: Solid Isotropic Material with Penalization (SIMP) and level-set.

<u>Solid Isotropic Material with Penalization (SIMP)</u>

In SIMP, pseudo-densities are assigned to finite-elements, and optimized to meet the desired objective [20]. The primary advantage of SIMP is that it is well-understood and relatively easy to implement [20]. Indeed, SIMP has been applied to almost all types of problems ranging from fluids to non-linear structural mechanics problems. However, the 'singularity-problem' associated with zero-density elements require careful treatment, for example through epsilon-methods [12], [21], [22]. Secondly, the ill-conditioning of the stiffness matrices, due to low-density elements, can lead to high computational costs for iterative solvers [16], [23].

One of the earliest implementation of SIMP for stress-constrained topology optimization was reported in [13], where the authors addressed instability and singularity issues via a weighted combination of compliance and global stress measure.

Since it is impossible to impose stress constraints at all points within the domain, element-stresses are typically lumped together into a single global quantity via the p-norm [24], Kreisselmeier–Steinhauser function [25], or potentially active constraints [26], and global/local penalization [27]. The equivalence of these two measures and their justification is discussed, for example, in [28]. Later in this paper, we shall exploit the p-norm global measure. Alternately, active-set methodologies have also been proposed where a finite number of elements with the highest stress states are chosen to be active during a given iteration [29], [30].

In [31], the authors proposed a framework to design the material distribution of functionally graded structures with a tailored Von Mises stress field. In [25], the authors studied the weight minimization problems with global or local stress constraints, in which the global stress constraints are defined by the Kreisselmeier–Steinhauser function. The mixed finite element method (FEM) was proposed for stress-constrained topology optimization, to alleviate the challenges posed by displacement-based FEM [32].

More recently, the authors of [33] proposed a *conservative* global stress measure, and the objective function was constructed using the relationship between mean compliance and von Mises stress; the authors used a SIMP-based mesh-independent framework. In [29] Drucker–Prager failure criterion is considered within the SIMP framework to handle materials with different tension and compression behaviors.

<u>Level-Set</u>

The second strategy for solving topology optimization problems relies on defining the evolving topology via a level-set. Since the domain is well-defined at all times, the singularity problem does not arise, and the stiffness matrices are typically well-conditioned; see [34] for a recent review and comparison of level-set based methods in structural topology optimization.

The authors of [35] proposed a level-set based stress-constrained topology optimization; a similar approach was explored in [36]. To address irregular, i.e., non-rectangular domains, an iso-parametric approach to solving the Hamilton-Jacobi equation was explored by the authors. In the level-set implementation of [27], a new global stress measure was proposed. In [12], [30], the authors combine the advantages of level-set with X-FEM for accurate shape and topology optimization. The active-set methodology with augmented Lagrangian is used to alleviate stress-concentrations. A topological level-set method for handling stress and displacement constraints in single-load problems was proposed in [19].

<u>Multi-Load Problems</u>

For multi-load problems, one can either adopt a worst-case approach or a weighted approach; these are not necessarily equivalent [37]. In the former, one arrives at a feasible but non-optimal solution. In the latter, the weights are subjective and difficult to establish *a priori*; the final topology will depend on the weights [20], [38], [39]. Additionally, due to convergence issues, application-specific methods have also been developed [40], [41]. For truss structures, an alternate approach based on the "envelope strain energy" was proposed in [42], but its advantages for continuum structures is not known.

In [41], [43], [44], for multi-load problems, the authors propose an alternate discrete variable approach for mass minimization while satisfying various performance constraints, such as deflections, stress, etc. This has the advantage of synthesizing a minimum-mass solution that can satisfy many performance requirements.

Multi-load problems are fairly common in compliant-mechanism design [8], [9], [45]–[47]. Specifically, one must solve (at least) two problems: (1) the primary problem involving the external load, and (2) an auxiliary problem with a unit load at the 'output' location. Further, multiple objectives must be met in the design of compliant mechanisms. These objectives are usually combined into a single weighted objective involving quantities such as the internal strain energy and mutual strain energy [8], [48]. In addition, constraints are typically imposed on displacements where the force is applied [46], [49]. These constraints are absorbed into the objective through penalization. In [50], the topological level-set was exploited to solve multi-load problems, but the weights were determined in an *ad hoc* fashion.

Commercial topology optimization systems such as Optistruct [51] solve multi-load problems through a weighted approach. However, since such systems are typically based on SIMP, i.e., a density-formulation, one may arrive at a disconnected topology in multi-load scenarios.

## 3. TECHNICAL BACKGROUND

The proposed topology optimization method is based on the concept of topological sensitivity that is reviewed next.

### 3.1 Topological Sensitivity

Topological sensitivity captures the first order impact of inserting a small circular hole within a domain on various quantities of interest. This concept has its roots in the influential paper by Eschenauer [52], and has later been extended and explored by numerous authors [53]–[57], including generalization to arbitrary features [58]–[60].

Consider again the problem illustrated earlier in Figure 1. Let the quantity of interest be $Q$ (example: compliance) that is dependent on the field $u$. Suppose a tiny hole is introduced, i.e., modifying the topology, as illustrated in Figure 3. The solution $u$ of static equilibrium equation calculated by finite element method and the quantity $Q$ will change. The topological sensitivity (aka topological derivative) is defined in 2-D as:

$$\mathcal{T}_Q(p) \equiv \lim_{r \to 0} \frac{Q(r) - Q}{\pi r^2} \quad (3.1)$$

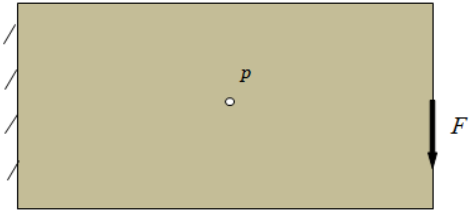

**Figure 3:** A topological change.

To find a closed-form expression for the topological sensitivity, one may first define an adjoint. Recall that the adjoint field associated with a quantity of interest satisfies [61]–[63]:

$$K\lambda = -\nabla_u Q \quad (3.2)$$

The right hand side of Equation (3.2) may be symbolically determined (see Section 4).

Once the adjoint is computed, the topological derivative is given by [64], [63]:

$$\mathcal{T}_Q = -\frac{4}{1+\nu}\sigma(u):\varepsilon(\lambda) + \frac{1-3\nu}{1-\nu^2}tr\big[\sigma(u)\big]tr\big[\varepsilon(\lambda)\big] \quad (3.3)$$

where

$$\begin{aligned}\sigma(u):&\ \text{Stress tensor of primary field}\\ \varepsilon(\lambda):&\ \text{Strain tensor of adjoint field}\end{aligned} \quad (3.4)$$

Thus, given the stress and strain field in the original domain (without the hole), one can compute the topological sensitivity over the entire domain.

Observe that, as a special case, when $Q = f^T u$, i.e., in the case of compliance, Equation (3.2) reduces to:

$$K\lambda = -f \quad (3.5)$$

In other words we arrive at $\lambda = -u$ as expected, and the topological sensitivity reduces to [64]:

$$\mathcal{T}_J(p) = \frac{4}{1+\nu}\sigma:\varepsilon - \frac{1-3\nu}{1-\nu^2}tr(\sigma)tr(\varepsilon) \quad (3.6)$$

If the domain is discretized into 2000 elements, the resulting field can be illustrated in Figure 4 in which the magnitude of topological sensitivity field is normalized to 1 and the elements where external forces are applied are manipulated with relatively high values to prevent singularities. In 3-D, the topological sensitivity field for compliance is given by [56]:

$$\mathcal{T}_J = -20\mu\sigma:\varepsilon + (2\mu - 3\lambda)tr(\sigma)tr(\varepsilon) \quad (3.7)$$

where $\mu$ and $\lambda$ are the Lame parameters.

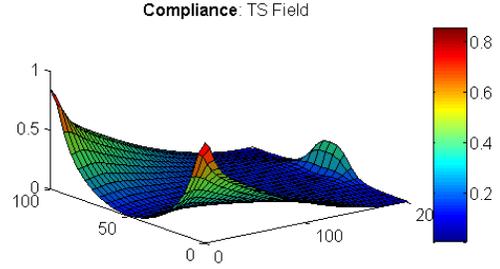

**Figure 4:** Topological sensitivity field.

### 3.2 Topological Level-set

A simple approach to exploiting topological sensitivity in topology optimization is to 'kill' mesh-elements with low values. However, this leads to instability and checker-board patterns. Alternately, the topological sensitivity field can be used to introduce holes during the topology optimization process via an auxiliary level-set [65]. Here, we directly exploit the topological sensitivity field as a level-set, as described next.

Consider again the compliance field illustrated in Figure 4; this is reproduced below in Figure 5a together with a cutting plane corresponding to an arbitrary cut-off value of $\tau = 0.03$. Given the field, and the cutting plane, one can define a domain $\Omega^\tau$ per:

$$\Omega^\tau = \{p \,|\, \mathcal{T}_J(p) > \tau\} \quad (3.8)$$

In other words, the domain $\Omega^\tau$ is the set of all points where the topological field exceeds $\tau$; the induced domain $\Omega^\tau$ is illustrated in Figure 5b. Now, the $\tau$ value can be chosen such that, say, 10% of the volume is removed. It is observed that the elements at the upper and lower corners on left end as well as where the force is applied have relatively high sensitivity values while the sensitivity values for the elements at corners on right end are relatively low. Since the elements with lower topological sensitivity values are least critical for the stiffness of the structure, they have more tendency to be eliminated. In other words, a 'pseudo-optimal' domain has been constructed directly from the topological sensitivity field.

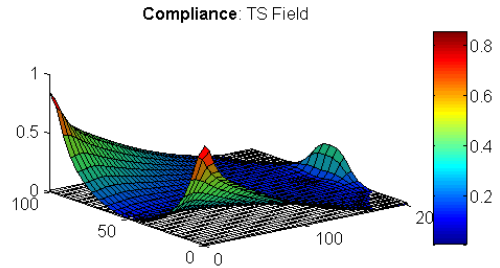

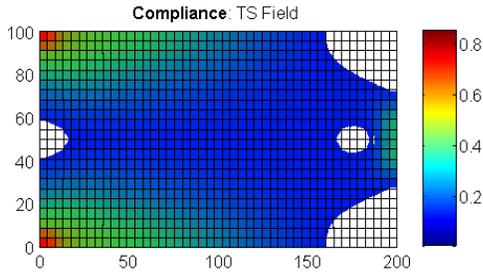

**Figure 5:** Compliance topological sensitivity, and the induced domain $\Omega^\tau$ for a volume fraction of 0.95.

However, the computed domain may not be 'optimal' [15], i.e., it may not be the stiffest structure for the given volume fraction. One must now repeat the following three steps: (1) solve the finite element problem over $\Omega^\tau$ (2) re-compute the topological sensitivity, and (3) find a new value of $\tau$ for the desired volume fraction. In essence, a fixed-point iteration is carried out [57], [66], [16], involving three quantities (see Figure 6): (1) domain $\Omega^\tau$, (2) displacement fields $u$ and $v$ over $\Omega^\tau$, and (3) topological sensitivity field over $\Omega^\tau$.

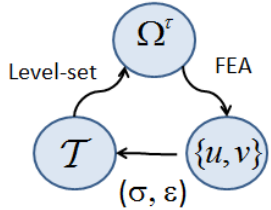

**Figure 6:** Fixed point iteration involving three quantities

Once convergence has been achieved (in typically 2~3 iterations), an optimal domain at 90% volume fraction will be obtained. An additional 10% volume can now be removed by repeating this process.

Using the above algorithm, the compliance problem posed in Equation (1.1) can be solved, resulting in a series of pareto-optimal topologies illustrated in Figure 7. Therefore, the algorithm finds pareto-optimal solutions to the problem:

$$\underset{\Omega \subset D}{Min}\{J, |\Omega|\} \tag{3.9}$$

Since all topologies are pareto-optimal, the constrained problem in Equation (1.3) is trivially solved by terminating the algorithm when the desired volume fraction has been reached.

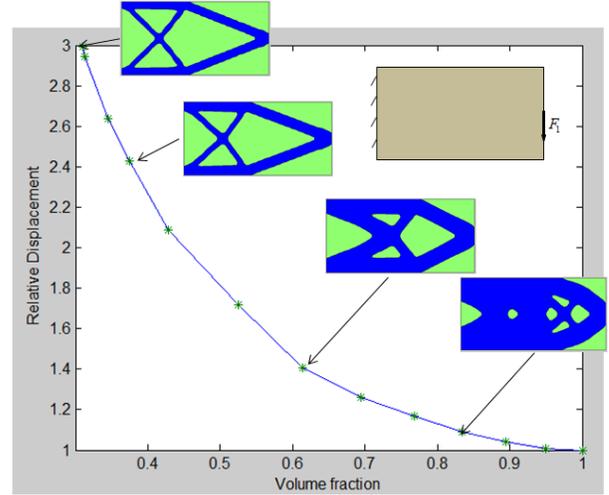

**Figure 7:** Pareto-optimal topologies

Observe that the above "PareTO" method is applicable to other objective functions (besides compliance) by replacing the compliance topological sensitivity field with the appropriate topological sensitivity field.

### 4. PROPOSED METHOD

The objective of this paper is to extend the above PareTO method to include arbitrary constraints, and multi-loads.

#### 4.1 Augmented Lagrangian Method

Towards this end, consider the classic *continuous-variable* constrained optimization problem:

$$\underset{x}{Min}\ f(x) \\ g_i(x) \leq 0 \tag{4.1}$$

Observe that this is a continuous variable problem involving a continuous variable $x$, as opposed to a topology optimization problem. One of the most popular methods for solving such problems is the *augmented Lagrangian method*, also referred to as the "Method of Multipliers" [19]. Since the augmented Lagrangian method is well established, we only provide a brief summary of the method.

In this method, the objective and the constraints are combined into a single unconstrained function, referred to as the augmented Lagrangian:

$$L(x, \mu, \gamma) = f(x) - \sum_{i=1}^{m} \bar{L}_i(x, \mu, \gamma) \tag{4.2}$$

In the above equation, $\tilde{L}_i(x, \mu, \gamma)$ is defined as [68]:

$$\bar{L}_i(x, \mu, \gamma) = \begin{cases} \mu_i g_i(x) - \frac{1}{2}\gamma_i(g_i(x))^2 & \mu_i - \gamma_i g_i(x) > 0 \\ \frac{1}{2}\mu_i^2 / \gamma_i & \mu_i - \gamma_i g_i(x) \leq 0 \end{cases} \tag{4.3}$$

where $\mu_i$ are the Lagrangian multipliers and $\gamma_i$ are the penalty parameters. The theory underlying the above definition is discussed, for example, in [68].

The Lagrangian multipliers and penalty parameters are initialized to an arbitrary set of positive values. Then, the Lagrangian in Equation (4.2) is minimized, typically via nonlinear conjugate gradient.

Towards this end, note that the gradient of the augmented Lagrangian is given by:

$$\nabla L(x,\mu,\gamma) = \nabla f - \sum_{i=1}^{m} \nabla \bar{L}_i(x,\mu,\gamma) \qquad (4.4)$$

where

$$\nabla \bar{L}_i(x,\mu,\gamma) = \begin{cases} \mu_i - \gamma_i g_i \ \nabla g_i & \mu_i - \sigma_i g_i(x) > 0 \\ 0 & \mu_i - \sigma_i g_i(x) \le 0 \end{cases} \qquad (4.5)$$

Once the minimization terminates, the Lagrangian multipliers are updated as follows [70]:

$$\mu_i^{k+1} = \max\{\mu_i^k - g_i(\hat{x}^k), 0\}, i = 1,2,3,...,m \qquad (4.6)$$

where the $\hat{x}^k$ is the minimum at the (current) $k$ iteration. The penalty parameters are also updated:

$$\gamma_i^{k+1} = \begin{cases} \gamma_i^k & \min(g_i^{k+1},0) \le \varsigma \min(g_i^k,0) \\ \max(\eta \gamma_i^k, k^2) & \min(g_i^{k+1},0) > \varsigma \min(g_i^k,0) \end{cases} \qquad (4.7)$$

where $0 < \varsigma < 1$ and $\eta > 0$; typically $\varsigma = 0.25$ and $\eta = 10$. The updates ensure rapid minimization of the objective, while satisfying the constraints.

The augmented Lagrangian is once again minimized and cycle is repeated until the objective cannot be reduced further. The implementation details and the robustness of the algorithm are discussed, for example, in [67], [68].

## 4.2 Augmented Topological Level-Set

Now consider the topology optimization problem:

$$\begin{aligned} & \underset{\Omega \subset D}{Min}\, \varphi \\ & g_i(u,\Omega) \le 0 \end{aligned} \qquad (4.8)$$

The goal is to extend the classic augmented Lagrangian method to solve the above problem. Drawing an analogy between Equations (4.1) and (4.8), we define the *topological augmented Lagrangian* as follows:

$$L(u,\Omega;\gamma_i,\mu_i) \equiv \varphi - \sum_{i=1}^{m} \bar{L}_i(u,\Omega;\gamma_i,\mu_i) \qquad (4.9)$$

where

$$\bar{L}_i(u,\Omega;\gamma_i,\mu_i) = \begin{cases} \mu_i g_i - \dfrac{1}{2}\gamma_i (g_i)^2 & \mu_i - \gamma_i g_i > 0 \\ \dfrac{1}{2}\mu_i^2 / \gamma_i & \mu_i - \gamma_i g_i \le 0 \end{cases} \qquad (4.10)$$

In classic continuous optimization, the gradient was defined with respect to the continuous variable $x$. Here, the gradient is defined with respect to a topological change. Drawing an analogy to the gradient operator in Equation (4.4), we propose the following *topological gradient operator*:

$$\mathcal{T}_\Omega[L(u,\Omega;\gamma_i,\mu_i)] \equiv \mathcal{T}_L = \mathcal{T}_\varphi - \sum_{i=1}^{m} \mathcal{T}_{\bar{L}_i} \qquad (4.11)$$

where $\mathcal{T}_\varphi$ is the topological level-set associated with the objective, and

$$\mathcal{T}_{\bar{L}_i} = \begin{cases} \mu_i - \gamma_i g_i \ \mathcal{T}_{g_i} & \mu_i - \gamma_i g_i > 0 \\ 0 & \mu_i - \gamma_i g_i \le 0 \end{cases} \qquad (4.12)$$

where

$$\mathcal{T}_{g_i} \equiv \mathcal{T}(g_i) \qquad (4.13)$$

are the topological level-sets associated with each of the constraint functions. Observe that we have essentially combined various topological level-sets into a single topological level set. The multipliers and penalty parameters are updated as described earlier.

The above concept easily generalizes to multi-load constrained topology optimization problem:

$$\begin{aligned} & \underset{\Omega \subset D}{Min}\, \varphi(u_1,u_2,..,u_N,\Omega) \\ & g_i(u_1,u_2,..,u_N,\Omega) \le 0;\ i = 1,2,...,I \end{aligned} \qquad (4.14)$$

in that the augmented Lagrangian is now defined as:

$$L \equiv \varphi - \sum_{i=1}^{m} \bar{L}_i(u_1,u_2,..,u_N,\Omega;\gamma_i,\mu_i) \qquad (4.15)$$

Thus, the only difference is that the constraint and objective depend on multiple displacement fields.

## 4.3 Illustrative Examples

Before we discuss implementation details, a few examples are provided to illustrate the concept of the augmented topological level-set.

### Displacement Constraint at a Point

Consider the single-load problem posed in Figure 8, where a y-displacement constraint is imposed at point $q$. The objective is to minimize volume fraction subject to a displacement constraint at a point.

$$\begin{aligned} & \underset{\Omega \subset D}{Min}\, |\Omega| \\ & u_y(q) - \delta_{max} \le 0 \end{aligned} \qquad (4.16)$$

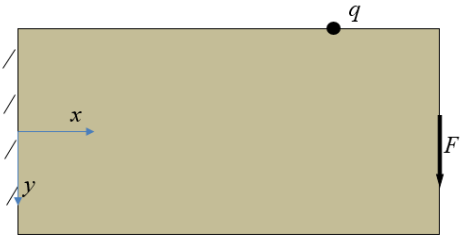

**Figure 8:** A single-load problem with displacement constraint.

First consider the objective function. It follows from Equation (3.1) that:

$$\mathcal{T}_\varphi \equiv \lim_{r \to 0} \frac{|\Omega_r| - |\Omega|}{\pi r^2} = \lim_{r \to 0} \frac{-\pi r^2}{\pi r^2} = -1 \qquad (4.17)$$

Next consider the displacement constraint. Since the point of interest does not coincide with the point of force-application, we first pose and solve an adjoint problem:

$$K\lambda = -\hat{\delta}_y(q) \qquad (4.18)$$

i.e., an auxiliary problem must be solved (see Figure 9).

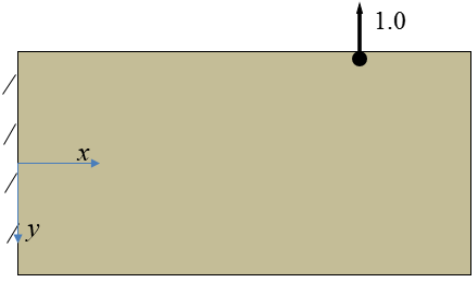

**Figure 9:** An auxiliary problem must be solved to obtain the adjoint.

Once the adjoint is obtained, the topological sensitivity of the constraint is obtained as usual via:

$$\mathcal{T}_g = -\frac{4}{1+\nu}\sigma(u):\varepsilon(\lambda) + \frac{1-3\nu}{1-\nu^2} tr\big[\sigma(u)\big] tr\big[\varepsilon(\lambda)\big] \quad (4.19)$$

Therefore, the combined topological level-set is given by:

$$\mathcal{T}_L = -1 - \mathcal{T}_{\bar{L}} \quad (4.20)$$

where

$$\mathcal{T}_{\bar{L}} = \begin{cases} \mu - \gamma g \ \mathcal{T}_g & \mu - \gamma g > 0 \\ 0 & \mu - \gamma g \leq 0 \end{cases} \quad (4.21)$$

Global p-norm Stress Constraint

Now consider a global stress constraint:

$$\underset{\Omega \subset D}{Min}|\Omega|$$
$$\sigma - \sigma_{\max} \leq 0 \quad (4.22)$$

where the global stress is defined by weighting the von Mises stresses over all elements via the popular p-norm:

$$\sigma = \left(\sum_e (\sigma_e)^p\right)^{1/p} \quad (4.23)$$

Computing the adjoint and the gradient of this global constraint is described in [18]. Once the adjoint has been computed, the topological level-set is defined as in Equation (4.19), followed by the augmented level-set as in Equation (4.20).

Multi-load Displacement Constraint

As an example of a multi-load problem, consider Figure 10, where the objective is to minimize volume such that the y-displacement at point *q* does not exceed a prescribed value under two different load conditions, i.e.,

$$\underset{\Omega \subset D}{Min}|\Omega|$$
$$u_{1y}(q) - \delta_{\max} \leq 0 \quad (4.24)$$
$$u_{2y}(q) - \delta_{\max} \leq 0$$

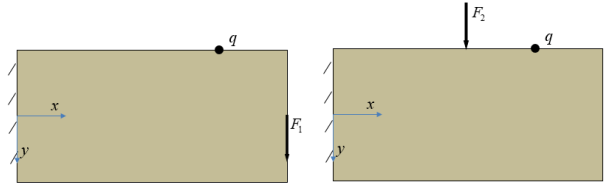

**Figure 10:** A multi-load problem with displacement constraint.

Three different topological sensitivity fields must be computed. As before, the field associated with the objective is:

$$\mathcal{T}_\varphi = -1 \quad (4.25)$$

Next, since the constraint is applied at point *q*, a unit load is used to construct a *single adjoint field* per Equation (4.18). Given the two displacements fields and the adjoint fields, the remaining two topological sensitivity fields are computed as follows:

$$\mathcal{T}_{g_1} = -\frac{4}{1+\nu}\sigma(u_1):\varepsilon(\lambda) + \frac{1-3\nu}{1-\nu^2} tr\big[\sigma(u_1)\big] tr\big[\varepsilon(\lambda)\big] \quad (4.26)$$

$$\mathcal{T}_{g_2} = -\frac{4}{1+\nu}\sigma(u_2):\varepsilon(\lambda) + \frac{1-3\nu}{1-\nu^2} tr\big[\sigma(u_2)\big] tr\big[\varepsilon(\lambda)\big] \quad (4.27)$$

**4.5 Proposed Algorithm**

The overall algorithm is illustrated in Figure 11, and described below.

1. The domain, desired volume fraction is initialized as described earlier. The volume decrement $\triangle v$ is initialized to be 2.5% of total volume fraction. As mentioned before, the Lagrangian multipliers $\mu_i$ and penalty parameters $\gamma_i$ can initialized to any arbitrary positive values. In this experiment, we set $\mu_i^0 = 1$ and $\gamma_i^0 = 10$.

2. Multiple FEAs are performed depending on the number of loads and adjoint problems.

3. The constraints are evaluated, and the multipliers and penalty parameters are updated.

4. If the constraints are satisfied proceed to step-5, else proceed to step-9.

5. The topological sensitivity fields for the objective and constraints are computed, and the augmented topological level-set is extracted.

6. The iso-surface for current volume fraction is extracted.

7. Check for the convergence of the topology. It is noted that if the relative compliance changes during the last 2 iterations are both smaller than 0.015, it is assumed this step is converged. If the topology has converged, then proceed to step-8, else return to step-2.

8. The next target-volume is decremented; if the desired volume has been reached the algorithm terminates, else it returns to step-2.

9. If the volume decrement is too small the algorithm terminates, else algorithm returns to step-2.

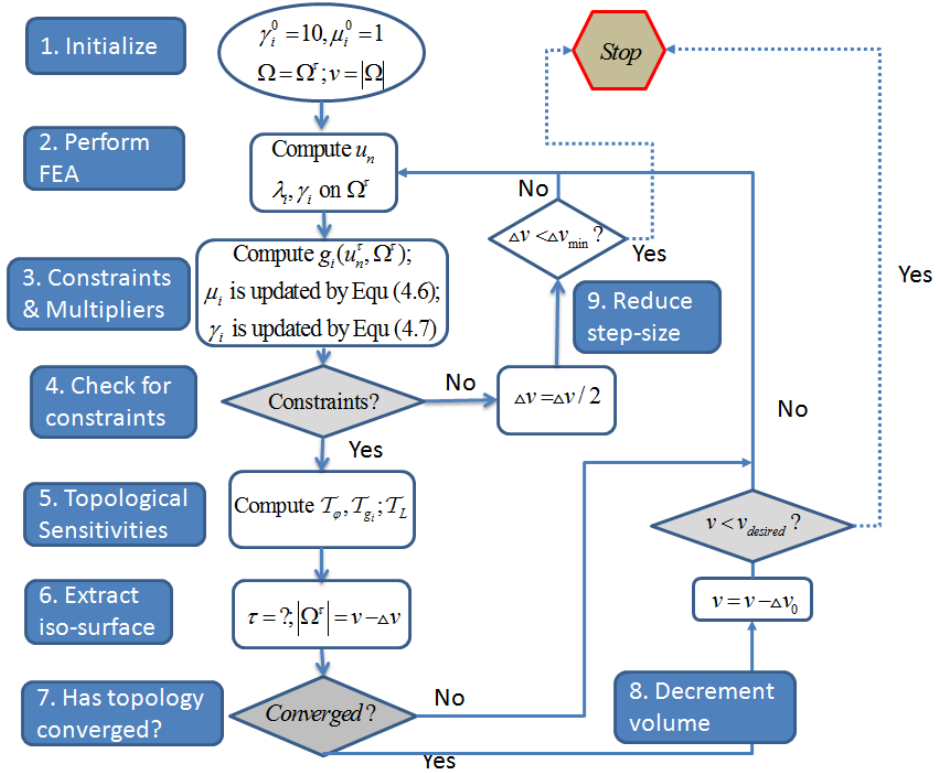

**Figure 11**: Proposed algorithm.

## 5. NUMERICAL EXAMPLES

In this Section, we demonstrate the proposed method through numerical experiments. The default material properties are $E = 2*10^{11}$ and $v = 0.33$. All experiments were conducted using Matlab 2013a on a Windows 7 64-bit machine with the following hardware: Intel I7 960 CPU quad-core running at 3.2GHz with 6 GB of memory.

Four-node quadrilateral finite elements are used in all experiments. All constraints are relative to the initial displacement and stresses, prior to optimization. Thus, a constraint:

$$u_y(q) - 3.0 \leq 0 \tag{5.1}$$

implies that the y-displacement at point q must not exceed three times the initial y-displacement at that point, prior to optimization. The constraint:

$$\sigma - 2.0 \leq 0 \tag{5.2}$$

implies that the maximum von Mises stress must not exceed twice the maximum von Mises stress prior to optimization.

### 5.1 L-bracket: Displacement & Stress Constraints

Before the multi-constrained examples, the effects of mesh size on final topologies are studied first. We start with a classic L-bracket problem (see Figure 12).

We consider a displacement constraint (at the point of force application) and a global stress constraint problem over the L-bracket and discretized the domain with five different mesh densities.

$$\underset{\Omega \subset D}{Min} |\Omega|$$
$$u_y - 1.5 \leq 0 \tag{5.3}$$
$$\sigma - 1000 \leq 0$$

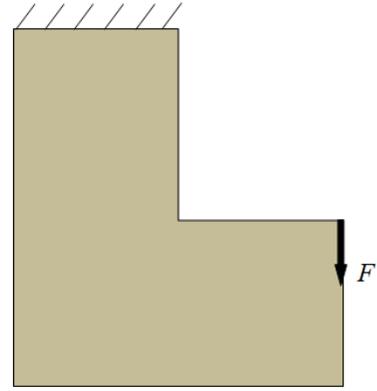

**Figure 12:** A single-load L-bracket problem.

Mesh densities, final topologies and their corresponding final volume fractions are shown in Table 1. It is observed that compared with mesh size 2000, the volume fraction of topology of mesh size 10000 is decreased by 4.0%.

**Table 1:** Results for problem in Figure 12.

| Mesh densities | Final results w/o filtering | Final results with filtering |
|---|---|---|
| 2000 | V=0.49 | V=0.31 |
| 4000 | V=0.48 | V=0.30 |
| 6000 | V=0.48 | V=0.31 |
| 8000 | V=0.47 | V=0.31 |
| 10000 | V=0.47 | V=0.31 |

Variation of relative compliance and maximum stress with respect to the mesh densities are shown in Figure 13. It is noted in Figure 13 that from mesh size 2000 to 10000, the relative compliances only change by 0.4%, while the relative stresses are changed by 61.5%.

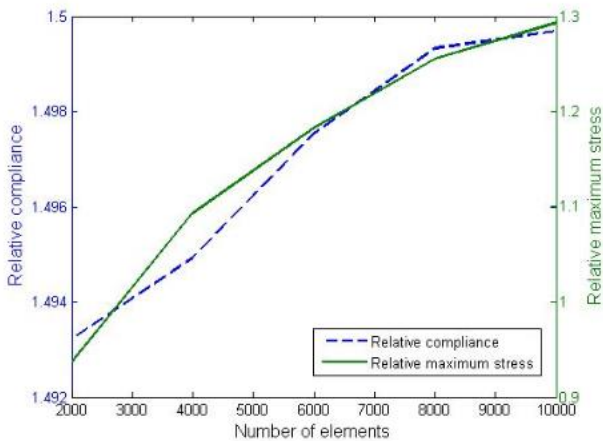

**Figure 13:** A comparison of relative compliance and maximum stress for different mesh sizes.

It is observed from Table 1 and Figure 13 that the final optimization results depend on the mesh density, i.e., with the increase of element numbers, the final volume fractions are slightly decreasing and maximum stress values see an increase while the compliance values almost stay constant.

In order to study the efficiency of this algorithm, a relationship between the FEA iteration numbers and condition numbers of corresponding assembled global stiffness matrix is shown in Figure 14.

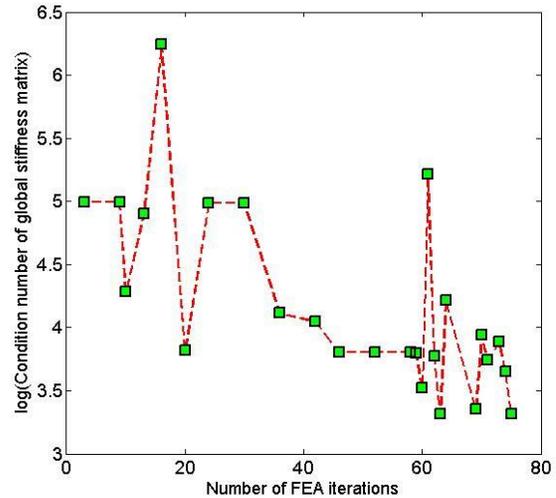

**Figure 14:** Condition number of global stiffness matrix in our proposed method.

It is noted that the condition number of global stiffness matrix is significantly decreased with the increase of iterations. Due to the "white-black" pattern and continuum structure generated by our PareTo method, the global stiffness matrix is well-behaved compared with SIMP in Figure 15.

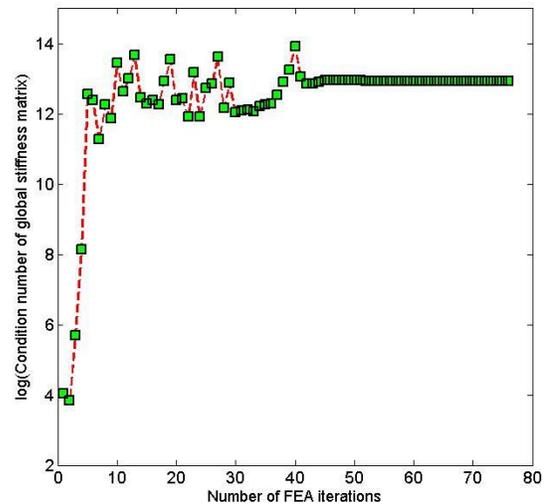

**Figure 15:** Change of condition number of global stiffness matrix in SIMP.

Figure 16 shows the tendency of the relative compliance and stress with respect to volume fractions during the optimization process for element number equaling 2000.

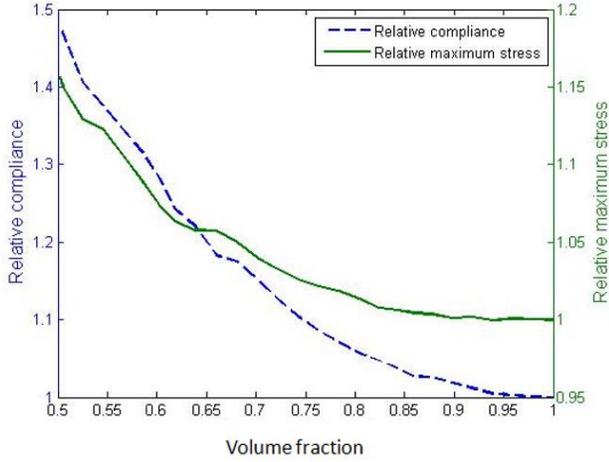

**Figure 16:** A relationship between relative compliance and volume fraction.

We now consider different combinations of displacement constraints and global stress constraints over the L-bracket:

$$\begin{aligned} &\underset{\Omega \subset D}{Min} |\Omega| \\ &u_y - \delta^{\max} \leq 0 \\ &\sigma - \sigma^{\max} \leq 0 \end{aligned} \quad (5.4)$$

The domain is discretized into 2000 finite elements:

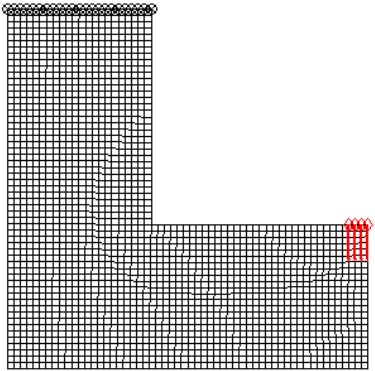

**Figure 17:** Finite element mesh for L-bracket.

The specific constraints and the final results are summarized in Table 2.

One can observe that, if the displacement constraint is active, then the topology corresponds to the classic 'compliance-minimization' problem, and if the stress constraint is active, the topology corresponds to the stress constraint problem [18].

**Table 2:** Constraints and results for problem in the Figure 12.

| Constraints | Volume fractions | Final displacements | Final topologies |
|---|---|---|---|
| $\delta^{\max} = 1000$<br>$\sigma^{\max} = 1.5$ | 0.34 | $\delta_J^{result} = 2.55$<br>$\boxed{\sigma^{result} = 1.50}$ | |
| $\delta^{\max} = 1.5$<br>$\sigma^{\max} = 1000$ | 0.49 | $\boxed{\delta^{result} = 1.50}$<br>$\sigma^{result} = 1.14$ | |
| $\delta^{\max} = 1.5$<br>$\sigma^{\max} = 1.1$ | 0.53 | $\boxed{\delta^{result} = 1.50}$<br>$\boxed{\sigma^{result} = 1.09}$ | |

It is observed from Figure 14 and Figure 15 that the finite element iteration numbers for the classic SIMP and PareTo methods are around 80; while the proposed algorithm takes up to 198 FEA iterations to solve a two-constraint optimization problem as listed in the first row of Table 2. Due to the nature of augmented Lagrangian method, the multipliers to constraints have to be updated in a sequential manner. It therefore may lead to high number of iterations. In a future study, we will include a comparison for the FEA iteration numbers between augmented Lagrangian implemented SIMP and our method.

**5.2 L-bracket: Multi-load, Multi-Constraint**

In this experiment, we consider the multi-load structure in Figure 18, where the topology optimization problem is

$$\begin{aligned} &\underset{\Omega \subset D}{Min} |\Omega| \\ &u_{1y} - \delta_1^{\max} \leq 0 \\ &\sigma_1 - \sigma_1^{\max} \leq 0 \\ &u_{2x} - \delta_2^{\max} \leq 0 \\ &\sigma_2 - \sigma_2^{\max} \leq 0 \end{aligned} \quad (5.5)$$

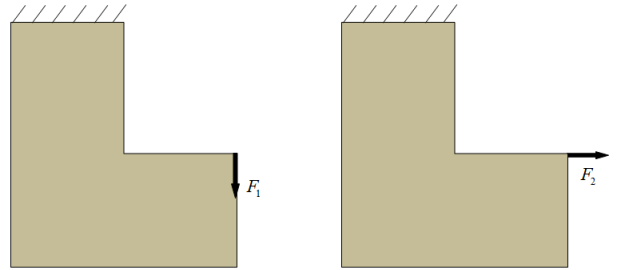

**Figure 18:** A multi-load L-bracket problem.

The results are summarized in Table 3; the final topologies depend strongly on the nature of the constraints.

**Table 3:** Constraints and results for problem in the Figure 18.

| Constraints | Final displacements | Final topologies |
|---|---|---|
| $\delta_1^{\max} = 1.5$<br>$\delta_2^{\max} = 10000$<br>$\sigma_1^{\max} = 10000$<br>$\sigma_2^{\max} = 10000$ | $\boxed{\delta_1^{result} = 1.50}$<br>$\delta_2^{result} = 5.37$<br>$\sigma_1^{result} = 1.17$<br>$\sigma_2^{result} = 2.80$ | 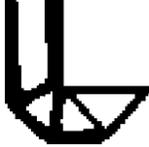<br>V=0.50 |
| $\delta_1^{\max} = 10000$<br>$\delta_2^{\max} = 1.5$<br>$\sigma_1^{\max} = 10000$<br>$\sigma_2^{\max} = 10000$ | $\delta_1^{result} = 26.44$<br>$\boxed{\delta_2^{result} = 1.50}$<br>$\sigma_1^{result} = 10.89$<br>$\sigma_2^{result} = 1.29$ | 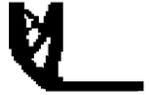<br>V=0.34 |
| $\delta_1^{\max} = 10000$<br>$\delta_2^{\max} = 10000$<br>$\sigma_1^{\max} = 1.5$<br>$\sigma_2^{\max} = 10000$ | $\delta_1^{result} = 2.47$<br>$\delta_2^{result} = 42.79$<br>$\boxed{\sigma_1^{result} = 1.50}$<br>$\sigma_2^{result} = 7.87$ | 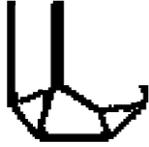<br>V=0.34 |
| $\delta_1^{\max} = 10000$<br>$\delta_2^{\max} = 10000$<br>$\sigma_1^{\max} = 10000$<br>$\sigma_2^{\max} = 1.5$ | $\delta_1^{result} = 68.01$<br>$\delta_2^{result} = 2.40$<br>$\sigma_1^{result} = 16.78$<br>$\boxed{\sigma_2^{result} = 1.50}$ | 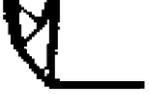<br>V=0.23 |
| $\delta_1^{\max} = 1.50$<br>$\delta_2^{\max} = 1.50$<br>$\sigma_1^{\max} = 1.50$<br>$\sigma_2^{\max} = 1.50$ | $\boxed{\delta_1^{result} = 1.50}$<br>$\boxed{\delta_2^{result} = 1.37}$<br>$\sigma_1^{result} = 1.20$<br>$\sigma_2^{result} = 1.16$ | 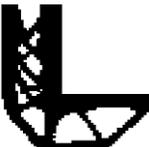<br>V=0.61 |

### 5.3 Cantilever Beam: Displacement Constraints

This experiment involves the classic 2-D cantilever beam illustrated in Figure 19. A point of interest 'q' is located in the middle of the top edge. The problem is:

$$\underset{\Omega \subset D}{Min} |\Omega|$$
$$u_y(q) - \delta_q^{\max} \leq 0 \quad (5.6)$$
$$u_y(a) - \delta_a^{\max} \leq 0$$

Thus, a displacement constraint is placed at the point-of-force application 'a', and a secondary point-of-interest 'q'.

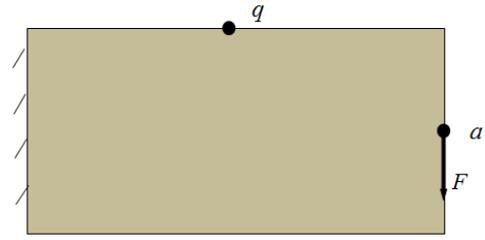

**Figure 19:** A single load cantilever beam problem.

Specific values for the allowable relative displacements at both points of interest are specified in Table 4. For FEA, the domain was discretized into 2000 elements, as illustrated in Figure 20.

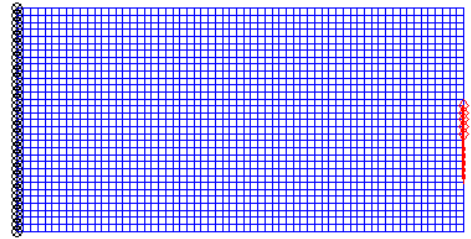

**Figure 20:** Finite element mesh for a cantilever beam.

The final volume fractions, the actual relative displacements reached, and the final topologies are also illustrated in Table 4.

The active constraints for each of the test cases is identified with 'box'; observe that, at least one of the constraints is active at termination.

**Table 4:** Constraints and results for problem in the Figure 11.

| Constraints | Volume fractions | Final displacements | Final topologies |
|---|---|---|---|
| $\delta_a^{\max} = 10.00$<br>$\delta_q^{\max} = 1.50$ | 0.48 | $\delta_a^{result} = 1.75$<br>$\boxed{\delta_q^{result} = 1.50}$ | 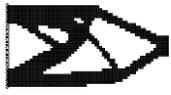 |
| $\delta_a^{\max} = 1.50$<br>$\delta_q^{\max} = 10.00$ | 0.55 | $\boxed{\delta_a^{result} = 1.50}$<br>$\delta_q^{result} = 1.63$ | 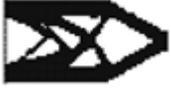 |
| $\delta_a^{\max} = 1.50$<br>$\delta_q^{\max} = 1.50$ | 0.56 | $\boxed{\delta_a^{result} = 1.50}$<br>$\delta_q^{result} = 1.40$ | 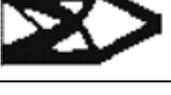 |

### 5.4 Cantilever Beam: Multi-load

We now consider a multi-load problem illustrated in Figure 21. The displacement constraint for each load is placed at the point of force application, i.e., the problem is:

$$\underset{\Omega \subset D}{Min} |\Omega|$$
$$u_{1y} - \delta_1^{\max} \leq 0 \quad (5.7)$$
$$u_{2x} - \delta_2^{\max} \leq 0$$

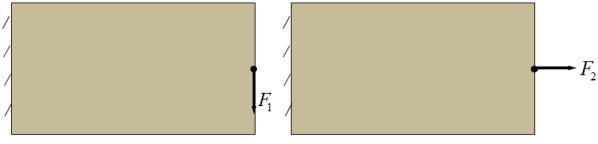

**Figure 21:** A multi-load cantilever beam problem.

The specific constraints and the final results are summarized in Table 5. Observe that the final topology is strongly dependent on the constraints.

**Table 5:** Constraints and results for problem in the Figure 21.

| Constraints | Volume fractions | Final displacements | Final topologies |
|---|---|---|---|
| $\delta_1^{\max}=1.50$ $\delta_2^{\max}=50.00$ | 0.59 | $\boxed{\delta_1^{result}=1.50}$ $\delta_2^{result}=1.47$ | |
| $\delta_1^{\max}=50.0$ $\delta_2^{\max}=1.50$ | 0.48 | $\delta_1^{result}=5.87$ $\boxed{\delta_2^{result}=1.50}$ | |
| $\delta_1^{\max}=1.50$ $\delta_2^{\max}=1.50$ | 0.62 | $\boxed{\delta_1^{result}=1.50}$ $\delta_2^{result}=1.36$ | |

### 5.5 Mitchell Bridge: Multi-load, Multi-Constraint

We now solve the multi-load, multi-constraint problem posed in Equation (5.5) over the classic Mitchell bridge structure in Figure 22. The domain is discretized into 2000 quadrilateral elements. The results are summarized in Table 6.

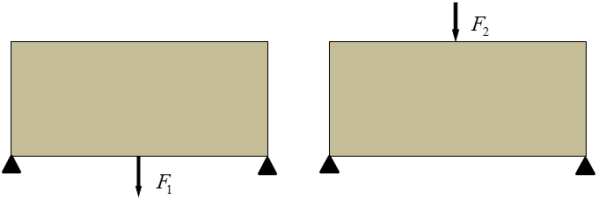

**Figure 22:** A multi-load Mitchell bridge problem.

**Table 6:** Constraints & results for problem in Figure 22.

| Constraints | Final displacements | Final topologies |
|---|---|---|
| $\delta_1^{\max}=1.50$ $\delta_2^{\max}=10.00$ $\sigma_1^{\max}=10.00$ $\sigma_2^{\max}=10.00$ | $\boxed{\delta_1^{result}=1.50}$ $\delta_2^{result}=1.32$ $\sigma_1^{result}=1.04$ $\sigma_2^{result}=1.03$ | V=0.51 |
| $\delta_1^{\max}=10.00$ $\delta_2^{\max}=1.50$ $\sigma_1^{\max}=10.00$ $\sigma_2^{\max}=10.00$ | $\delta_1^{result}=2.77$ $\boxed{\delta_2^{result}=1.50}$ $\sigma_1^{result}=1.89$ $\sigma_2^{result}=1.09$ | V=0.40 |
| $\delta_1^{\max}=10.00$ $\delta_2^{\max}=10.00$ $\sigma_1^{\max}=1.50$ $\sigma_2^{\max}=10.00$ | $\delta_1^{result}=4.12$ $\delta_2^{result}=3.18$ $\boxed{\sigma_1^{result}=1.50}$ $\sigma_2^{result}=1.22$ | 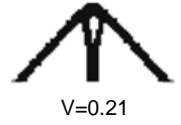 V=0.21 |
| $\delta_1^{\max}=10.00$ $\delta_2^{\max}=10.00$ $\sigma_1^{\max}=10.00$ $\sigma_2^{\max}=1.50$ | $\delta_1^{result}=5.68$ $\delta_2^{result}=4.15$ $\sigma_1^{result}=2.05$ $\boxed{\sigma_2^{result}=1.47}$ | 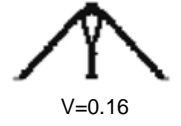 V=0.16 |
| $\delta_1^{\max}=1.50$ $\delta_2^{\max}=1.50$ $\sigma_1^{\max}=1.50$ $\sigma_2^{\max}=1.50$ | $\boxed{\delta_1^{result}=1.50}$ $\delta_2^{result}=1.36$ $\sigma_1^{result}=1.03$ $\sigma_2^{result}=1.01$ | 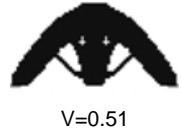 V=0.51 |

### 6. CONCLUSIONS

The main contribution of the paper is a new method for multi-load, multi-constrained topology optimization, where the topological sensitivity field for each loading and each constraint is computed, and then combined via augmented Lagrangian methods. This is then exploited to generate a set of pareto-optimal topologies. As illustrated via numerical examples, the proposed not only generates topologies consistent with those published in the literature, but provides solutions to more challenging problems that have not been considered before.

### Acknowledgements

The authors would like to thank the support of National Science Foundation through grants CMMI-1232508 and CMMI-1161474.